\documentclass[12pt]{iopart}

\usepackage{graphicx,color}
\usepackage{dcolumn}
\newcolumntype{.}{D{.}{.}{-1}}
\usepackage{bm}
\usepackage{epstopdf}
\usepackage{subfigure}
\usepackage{textcomp}

\newcommand{\cu}{$\alpha$-Cu$_2$V$_2$O$_7$}

\newcommand{\al}{$\alpha$}

\newcommand{\tc}{$T_{\rm C}$}
\newcommand{\tp}{$T^{\prime}$}

\newcommand{\mb}{\(\mu _{\rm B}\)}
\newcommand{\mbfu}{\(\mu _{\rm B}/{\rm f.u.}\)}
\newcommand{\tn}{$T_{\rm N}$}
\newcommand{\ts}{$T_{\rm s}$}

\newcommand{\cp}{$c_{\rm p}$}

\newcommand{\jmk}{J/(mol\,K)}

\newcommand{\bc}{$B_{\rm C}$}

\newcommand{\bres} {$B_{\rm{res}}$}

\begin{document}

\title[Magnetoelastic coupling and spin excitations in $\alpha$-Cu$_2$V$_2$O$_7$]{Magnetoelastic coupling and ferromagnetic-type in-gap spin excitations in multiferroic $\alpha$-Cu$_2$V$_2$O$_7$}

\author{L.~Wang}
\address{Kirchhoff Institute of Physics, Heidelberg University, INF 227, D-69120 Heidelberg, Germany}
\address{Leibniz-Institute for Solid State and Materials Research (IFW) Dresden, 01069 Dresden, Germany}
\author{J.~Werner}
\address{Kirchhoff Institute of Physics, Heidelberg University, INF 227, D-69120 Heidelberg, Germany}
\author{A.~Ottmann}
\address{Kirchhoff Institute of Physics, Heidelberg University, INF 227, D-69120 Heidelberg, Germany}
\author{R.~Weis}
\address{Kirchhoff Institute of Physics, Heidelberg University, INF 227, D-69120 Heidelberg, Germany}
\author{M.~Abdel-Hafiez}
\address{Physics Department, Faculty of Science, Fayoum University, Fayoum 63514, Egypt}
\address{Physikalisches Institut, Goethe-Universit\"{a}t, D-60323 Frankfurt a.M., Germany}
\author{J. Sannigrahi}
\address{ISIS Facility, Rutherford Appleton Laboratory, Didcot, United Kingdom}
\author{S. Majumdar}
\address{Department of Solid State Physics, Indian Association for the Cultivation of Science, 2A \& B Raja S. C. Mullick Road, Jadavpur, Kolkata 700 032, India}
\author{C.~Koo}
\address{Kirchhoff Institute of Physics, Heidelberg University, INF 227, D-69120 Heidelberg, Germany}
\author{R.~Klingeler}
\ead{klingeler@kip.uni-hd.de}
\address{Kirchhoff Institute of Physics, Heidelberg University, INF 227, D-69120 Heidelberg, Germany}
\address{Centre for Advanced Materials (CAM), Heidelberg University, INF 225, D-69120 Heidelberg, Germany}


\vspace{10pt}
\begin{indented}
\item[]March 2018
\end{indented}

\begin{abstract}
We investigate magnetoelectric coupling and low-energy magnetic excitations in multiferroic $\alpha$-Cu$_2$V$_2$O$_7$ by detailed thermal expansion, magnetostriction, specific heat and magnetization measurements in magnetic fields up to 15~T and by high-field/high-frequency electron spin resonance studies. Our data show negative thermal expansion in the temperature range $\leq 200$~K under study. Well-developed anomalies associated with the onset of multiferroic order (canted antiferromagnetism with a significant magnetic moment and ferroelectricity) imply pronounced coupling to the structure. We detect anomalous entropy changes in the temperature regime up to $\sim 80$~K which significantly exceed the spin entropy. Failure of Gr\"{u}neisen scaling further confirms that several dominant ordering phenomena are concomitantly driving the multiferroic order. By applying external magnetic fields, anomalies in the thermal expansion and in the magnetization are separated. Noteworthy, the data clearly imply the development of a canted magnetic moment at temperatures above the structural anomaly. Low-field magnetostriction supports the scenario of exchange-striction driven multiferroicity. We observe low-energy magnetic excitations well below the antiferromagnetic gap, i.e., a ferromagnetic-type resonance branch associated with the canted magnetic moment arising from Dzyaloshinsii-Moriya interactions. The anisotropy parameter $\tilde{D}=1.6(1)$~meV indicates a sizeable ratio of DM- and isotropic magnetic exchange.
\end{abstract}


\section{Introduction}

Elucidating the mechanisms of multiferroicity and pushing the magnetoelectric coupling towards higher values are among the main challenges of current condensed matter physics. Despite the great potential for applications, there are only few materials where ferromagnetic and ferroelectric order coexist and hence offer the potential of mutually switching the magnetization and the electrical polarization by $E$- and $B$-fields, respectively.~\cite{Fiebig2005,Wang2009,Brink2008} One promising route to realise materials with considerable magnetoelectric coupling is to exploit unusual long-periodic spin ordered structures evolving in quasi-low-dimensional magnetic systems.~\cite{Cheong2007} The recent discovery of giant ferroelectric polarization and large magnetoelectric coupling in the magnetically ordered phase of \cu\ somehow reaffirms this general concept as the system may be described by spin-1/2 zig-zag chains with strong interchain coupling.~\cite{lee2016magnetism,sanchezJAP2011} While the chains consist of edge-sharing distorted CuO$_5$-polyhedra, the non-centrosymmetric orthorhombic $Fdd2$ structure of the $\alpha$-phase permits stronger interchain interaction than the other polymorphs of Cu$_2$V$_2$O$_7$.~\cite{alphaphase,Betapahase,gammaphase} Magnetism in \cu\ is rather three-dimensional as inelastic neutron studies suggest dominant interchain exchange interaction $J_3$ between third nearest neighbours, in addition to the nearest- and next-nearest neighbour interactions $J_1$ and $J_2$.~\cite{banerjee2016spin} Notably, long-range antiferromagnetic order evolving below \tc\ = 35~K exhibits a considerable magnetic moment arising from spin canting due to antisymmetric Dzyaloshinsky-Moriya (DM) interactions.~\cite{pommer2003interplay,Ponomarenko2000, gitgeatpong2015magnetic} It is associated with the simultaneous development of spontaneous electric polarization.~\cite{SannigrahiPRB2015}

Giant ferroelectric polarization in \cu\ is suggested to be induced by a symmetric exchange-striction mechanism, which indicates an improper nature of multiferroicity. It may be expected that, similar to other low-dimensional chain materials (e.g., Cu$_3$(CO$_3$)$_2$(OH)$_2$ \cite{CongPRB2014}) or other multiferroic materials (e.g., TbFe$_3$(BO$_3$)$_4$~\cite{HamannPRB2010}), that there are pronounced magnetoelastic effects in \cu . However, except for the observation of unusual negative thermal expansion~\cite{ZhangNTE} above the room temperature, neither dilatometric studies nor any magneto-structural investigations have been reported for \cu\ or another member of this class of materials. Our present study of thermal expansion and magnetostriction on \cu\ elucidates the interrelation of structural, magnetic, and electron degrees of freedom in this material. In particular, we investigate in detail the lattice distortions associated with the evolution of multiferroic order in \cu\ as well as the influence of external magnetic fields. A detailed magnetic phase diagram is mapped out which differs from the one reported recently in Ref.~\cite{gitgeatpong2017}. In addition to the analysis of the thermodynamic properties, we show the low-energy $q=0$ collective ferromagnetic mode detected by high-frequency electron spin resonance. Quantitatively, our analysis yields a large value of the effective anisotropy parameter $\tilde{D} = 1.6(1)$~meV.

\section{Experimental}

Polycrystalline \cu\ was prepared by conventional solid state synthesis as reported in Ref.~\cite{SannigrahiPRB2015}. Static magnetization $\chi=M/B$ was studied in magnetic fields up to 15~T by means of a home-built vibrating sample magnetometer~\cite{vsm} (VSM) and in fields up to 5~T in a Quantum Design MPMS-XL5 SQUID magnetometer. Specific heat measurements at 0~T and 9~T have been done in a Quantum Design PPMS using a relaxation method. The relative length changes $dL/L$ were studied on a cuboidal-shaped pressed pellet whose dimension in the measurement direction is 3.28~mm. The measurements were done by means of a three-terminal high-resolution capacitance dilatometer.~\cite{adem2010} In order to investigate the effect of magnetic fields, the thermal expansion coefficient $\alpha = 1/L\cdot dL(T)/dT$ was studied in magnetic fields up to 15~T. In addition, the field induced length changes $dL(B)/L$ were measured at various fixed temperatures in magnetic fields up to 15~T and the longitudinal magnetostriction coefficient $\lambda = 1/L\cdot dL(B)/dB$ was derived. The magnetic field was applied along the direction of the measured length changes. High-frequency electron spin resonance (HF-ESR) measurements were carried out using a phase-sensitive millimeter-wave vector network analyser (MVNA) from AB Millimetr\'{e} in the frequency range from 30 to 350 GHz.~\cite{Comba2015} For the experiments in magnetic fields up to 16~T, the cuboidal pressed pellet was placed in the sample space of the cylindrical waveguide.

\section{Results}

\subsection{Thermal expansion and specific heat at B=0}

\begin{figure}
\includegraphics[width=0.85\columnwidth,clip] {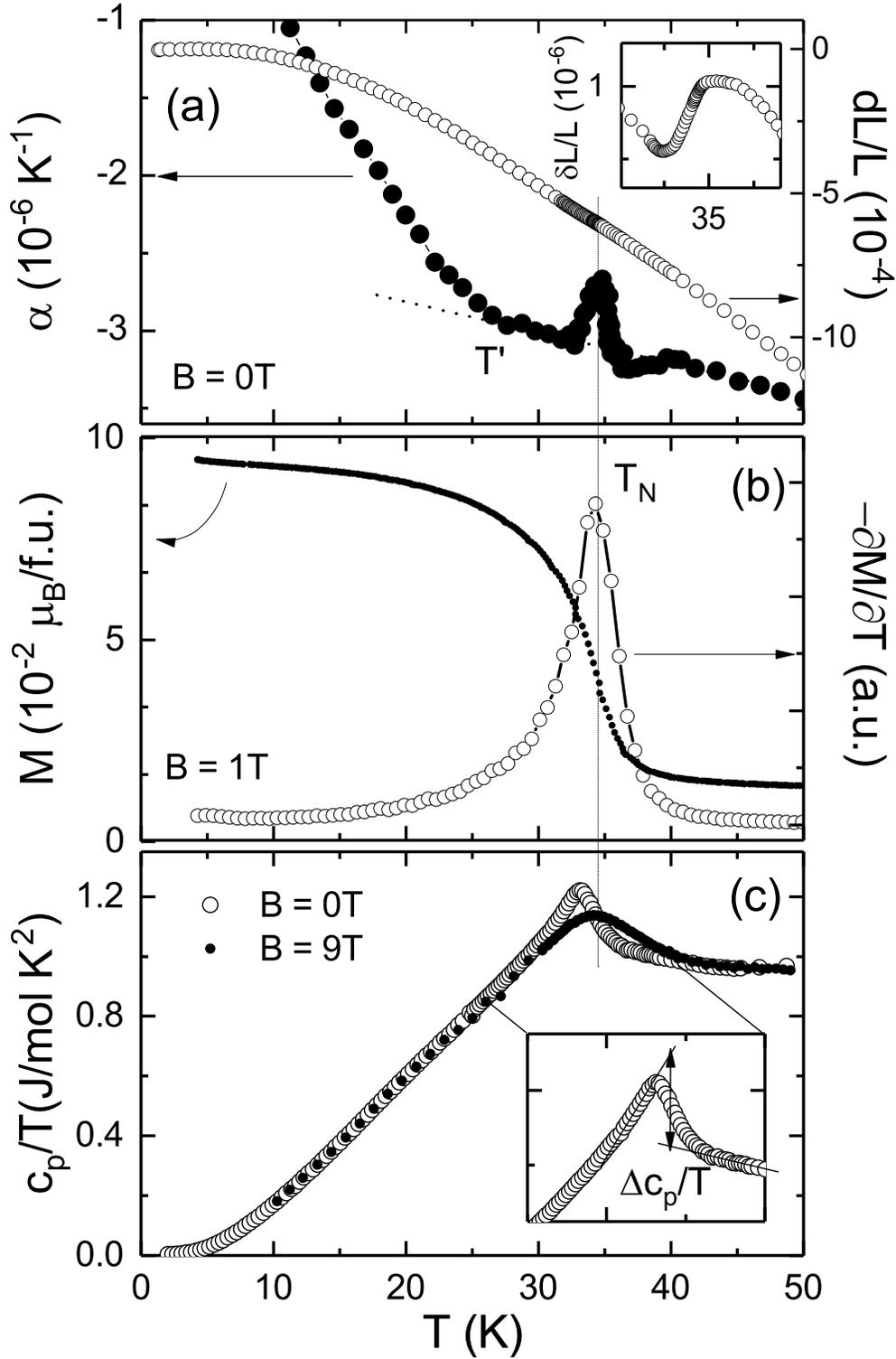}
\caption{(a) Length changes $dL/L$ and thermal expansion coefficient $\alpha$, (b) magnetization and its derivative $\partial M/\partial T$ measured at $B=1$~T, and (c) specific heat \cp\ at $B=0$ and 9~T. The dashed line shows \tn . \tp\ marks a kink in the thermal expansion coefficient. The inset in (a) shows $\delta L/L$ which is $dL/L$ at the anomaly after subtracting an arbitrary linear background fitted to the data outside the anomaly. The inset in (c) shows how the specific heat anomaly $\Delta c_{\rm p}$ has been obtained.}\label{fig:1T}
\end{figure}

Low-temperature thermal expansion of \cu\ is negative as illustrated by the temperature dependence of the length changes $dL/L$ and the thermal expansion coefficient \al\ in Fig.~\ref{fig:1T}a. This holds not only for T $<$ 50 K as shown in Fig.~\ref{fig:1T}a but for the whole temperature range up to 200~K under study (not shown) and hence is consistent with observation of negative thermal expansion in the temperature range from 300~K to about 550~K in a recent powder-XRD study~\cite{ZhangNTE}. In addition, there are pronounced lattice changes at \tn\ which show up in a peak-shaped anomaly of the thermal expansion coefficient $\alpha$ indicating \tn\ = 34$\pm$1~K. Concomitantly, the magnetization implies the formation of a significant spontaneous magnetic moment of about 0.1~\mb . As it was reported previously, the ferromagnetic-like response of our polycrstalline sample is associated with a weak spontaneous moment appearing for $B\|c$ only.~\cite{gitgeatpong2017} The derivative of magnetization $\partial M/\partial T$ (at $B=1$~T) shown in Fig.~\ref{fig:1T} qualitatively illustrates the evolution of the canted AFM phase and its anomaly at \tn\ resembles the one observed in \al . Although a weak discontinuous character of the transition, at \tn , is demonstrated by hysteresis effects reported in Ref.~\cite{SannigrahiPRB2015}, the evolution of the magnetization and of the length exhibits only very weak first order character but suggests a predominantly continuous behaviour. The experimentally measured specific heat anomaly in Fig.~\ref{fig:1T}c neither shows a peak-like nor a pronounced $\lambda$-shape anomaly but a rather step-like behaviour. Note, however, that a weakly discontinuous character of the anomaly may be smeared out by the applied calorimetric relaxation method. The broad anomaly in \cp\ appears rather jump-like, with the transition temperature at half of the specific heat jump $\Delta c_p\approx$ 9.3~\jmk .

The thermal expansion data imply that, at the magnetic transition, the volume of the unit cell shrinks when the canted antiferromagnetic phase evolves upon cooling. The signs of the anomalies in $\alpha$ and $dL/L$ hence imply a positive hydrostatic pressure effect on \tn , i.e. $\mathrm{d}T_\mathrm{N}/\mathrm{d}p > 0$. In order to further evaluate the anomalous length changes, in a phenomenological approach we have fitted $\alpha(T)$ well above and below \tn\ linearly as indicated in Fig.~\ref{fig:LvsT} and have subtracted this as a background. Integrating the resulting anomaly provides the anomalous length changes associated with the multiferroic transition, i.e., $\Delta L/L\approx 1.3 \times 10^{-6}$ at $B=0$~T. In addition to the anomaly at \tn , $\alpha(T)$ displays further features: (1) There is a regime of anomalous length changes above \tn\ extending up to around 50~K. (2) While $\alpha$ is only weakly temperature dependent at 27~K~$\leq T\leq$~32~K, there is a kink followed by a pronounced linear increase of $\alpha$ upon further cooling. At the kink temperatures \tp , there are no clear anomalies in \cp\ or $\partial M/\partial T$.

\subsection{Effect of external magnetic field}

\begin{figure}
\includegraphics[width=1.0\columnwidth,clip] {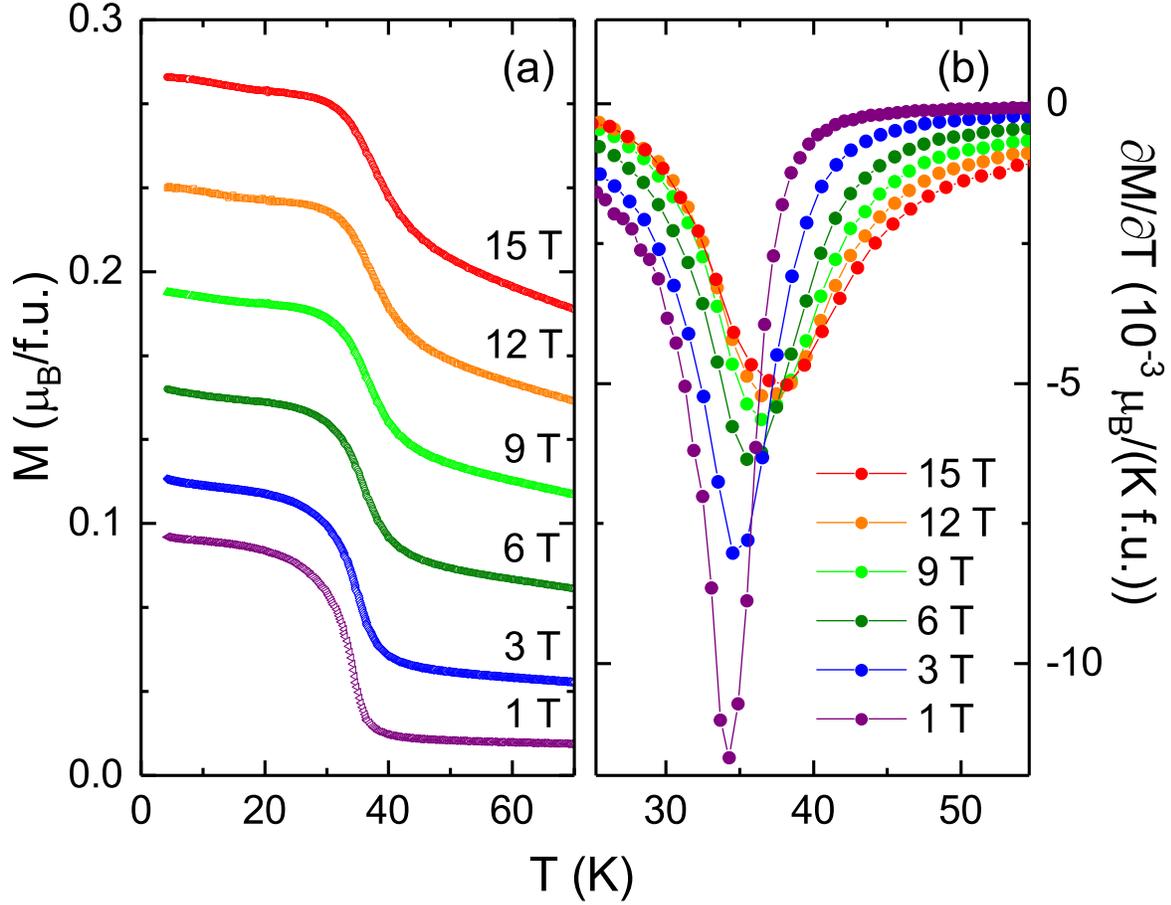}
\caption{(a) Magnetization and (b) the derivative $\partial M/\partial T$ vs. temperature in the vicinity of \tn\ at different magnetic fields. }\label{fig:MvsT}
\end{figure} 

\begin{figure}
\includegraphics[width=1.0\columnwidth,clip] {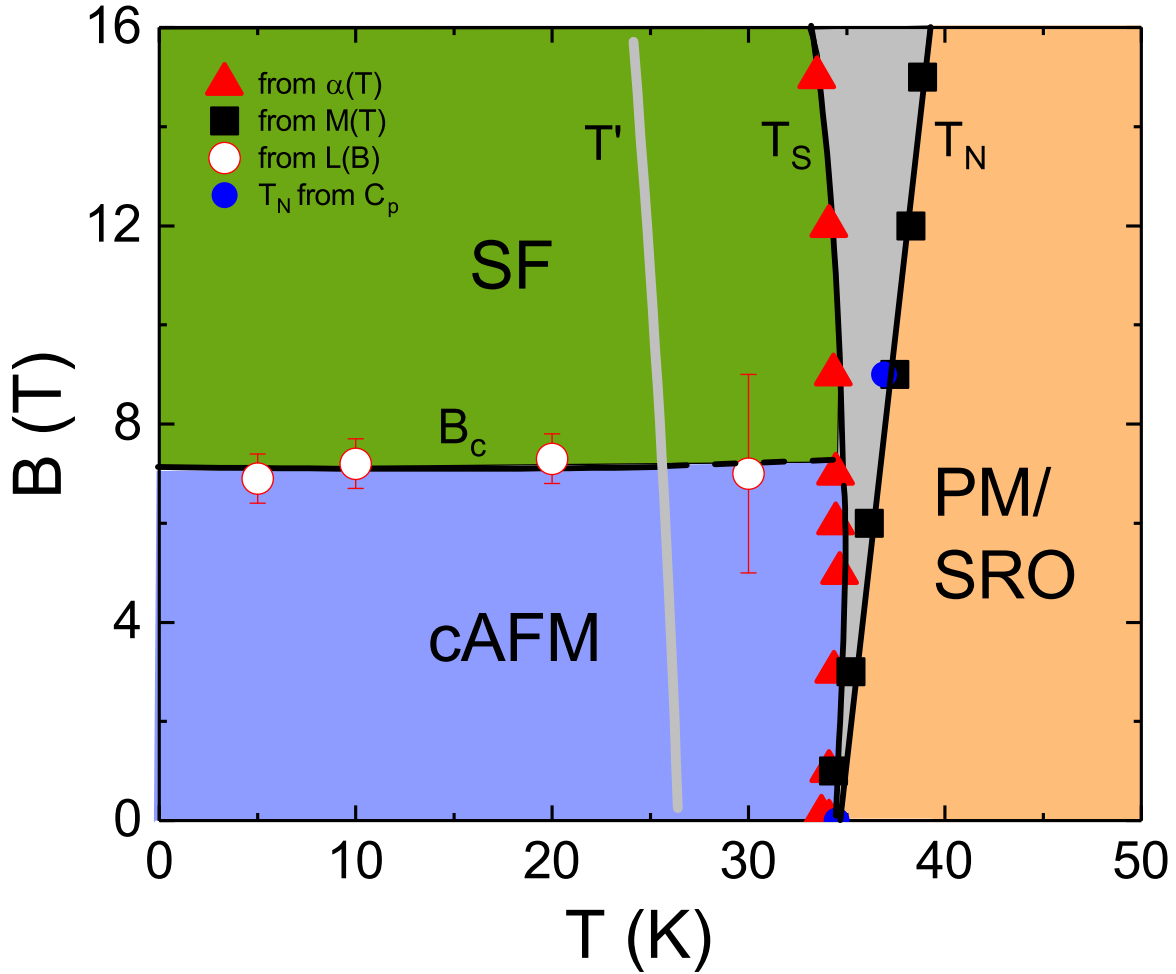}
\caption{Magnetic phase diagram of \cu\ as constructed from thermal expansion, longitudinal magnetostriction, specific heat, and magnetization measurements. The lines are guides to the eyes. cAFM and SF denote canted and spin-flopped AFM phases, PM/SRO means paramagnetic/short range ordered. \tn , \ts , and \bc\ denote the associated anomaly temperatures and fields. \tp\ shows the temperature of the kink in the thermal expansion coefficient.}\label{fig:PD}
\end{figure}

Application of external magnetic fields affects both the size and the temperature of the anomalies. A clear increase of $M$ at the phase transition is observed for all magnetic fields up to 15~T. The fact that, at \tn , \cu\ shows an increase of the magnetization due to the evolution of a canted antiferromagnetic phase (see Fig.~\ref{fig:MvsT}a) implies a positive field dependence of the phase boundary in the whole magnetic field range under study. This is indeed confirmed by the experimental data in Fig.~\ref{fig:MvsT}b where the minima in $\partial M/\partial T$ at different magnetic fields enable to deduce the boundary of the associated phase transition. The data show a positive magnetic field dependence $dT_{\rm N}/dB$ even in high magnetic fields up to $B=15$~T as displayed in the phase diagram in Fig.~\ref{fig:PD}. Quantitatively, the anomaly size does not change significantly but only very slightly decreases and broadens upon variation of $B$. The fact that \tn\ is associated with an increase of the magnetisation even at high magnetic fields thermodynamically implies the observed positive field dependence up to 15~T. Considering a temperature hysteresis found at \tn\ [\cite{SannigrahiPRB2015}], for a quantitative analysis we tentatively approximate the anomaly by a jump $\Delta M$ rather than a kink. This is consistent with the data in particular at higher magnetic fields where the anomaly seems to show a slightly discontinuous character. Evaluating the data correspondingly in the magnetic field range $9~{\rm T}\leq B\leq 15~{\rm T}$ yields a rather field independent jump $\Delta M = 0.054~\mu_{\rm B}$/f.u., at \tn ($B$). According to the Clausius-Clapeyron relation (e.g., Ref.~\cite{Stockert2012}), from the slope of the phase boundary \tn ($B$) we deduce that the onset of canted AFM order, at $B=15$~T, is associated with entropy changes of $\Delta S\approx 1.0(3)$~\jmk .


\begin{figure}[t]
\includegraphics[width=0.85\columnwidth,clip] {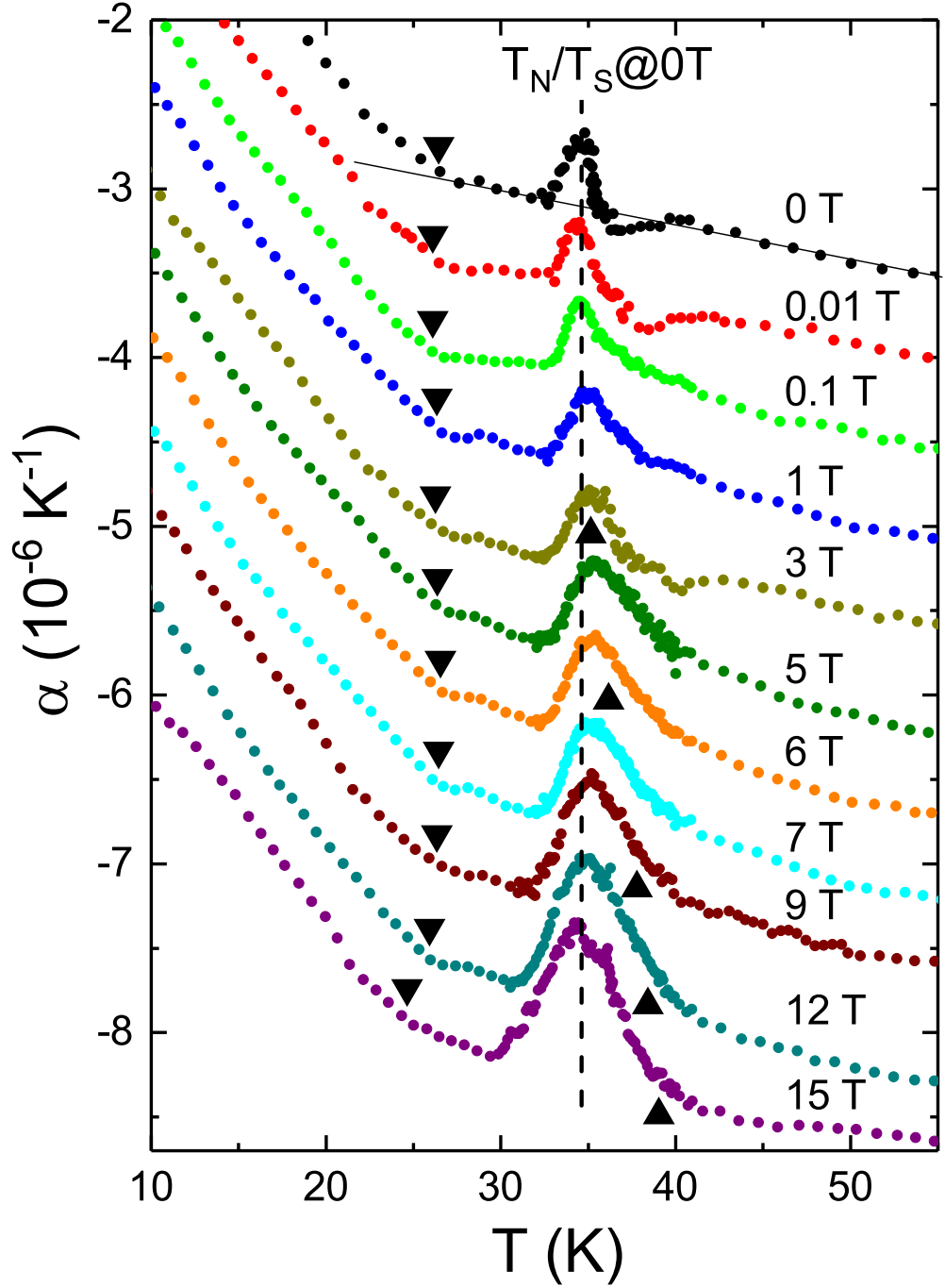}
\caption{Thermal expansion coefficient \al\ at various magnetic fields. The solid line is a guide to the eye to illustrate how \tp\ (black down triangles) is deduced. The vertical dashed line marks \tn\ = \ts\ at $B=0$~T. Black up triangles mark \tn ($B$) as derived from the minima in $dM/dT$ (Fig.~\ref{fig:MvsT}).}\label{fig:LvsT}
\end{figure}

The nature of the phase transition is further illustrated by the magnetic field dependence of the anomaly in the thermal expansion coefficient (Fig.~\ref{fig:LvsT}), which we label \ts . At $B=0$, \ts\ = \tn\ (see Fig.~\ref{fig:1T}). However, as compared to the anomaly in $\partial M/\partial B$, the peak maxima show a much smaller (but also positive) shift $dT_{\rm s}/dB>0$ for $B\leq 6$~T. The different field dependencies of \ts\ and \tn\ are unambiguously demonstrated by the behaviour at $B>6$~T where \ts\ is shifted to lower temperatures (Fig.~\ref{fig:LvsT}). In this field range, we hence find a $negative$ slope of the phase boundary, i.e., $dT_{\rm s}/dB<0$. In particular, upon application of external magnetic fields the peak in \al\ appears at lower temperature as compared to the anomaly in the magnetization.

To summarize, there is an anomaly at \tn\ which in the magnetization data is mainly signaled by the evolution of a small ferromagnetic moment but not associated with a clear hydrostatic pressure dependence at high magnetic fields. While, at \ts , there is an anomaly in the thermal expansion associated with only very small magnetization changes and corresponding small field dependence of the phase boundary. Both phase boundaries are shown in the magnetic phase diagram in Fig.~\ref{fig:PD}. We recall the fact that a small ferromagnetic moment appears for $B\|c$ only while no significant magnetic moment evolves for $B\perp c$.~\cite{gitgeatpong2017} This straightforwardly implies a significantly anisotropic magnetic field effect. The transition line \tn ($B$) in our phase diagram hence presumingly illustrates the effect of $B\parallel c$ which has not been reported in the literature yet. On the other hand, absence of a sizable magnetization anomaly for $B\perp c$ suggests insignificant field dependence of \tn ($B\perp c$). This suggests to associate \ts ($B$) to the effect of $B\perp c$. Indeed, \ts ($B$) roughly reproduces \tn ($B\parallel a$) from Ref.~\cite{gitgeatpong2017} for magnetic fields below 5 T which strongly supports this scenario that the two features represent the effect of magnetic field applied along different crystallographic directions. Note, however, that we cannot definitely exclude the presence of an intermediate phase at \ts\ $<T<$ \tn .

Remarkably, irrespective of the interpretation of the anomaly at \ts ($B$), the data imply that the bare evolution of the spontaneous magnetic moment at \tn ($B$) is \textsl{not} associated with clear volume changes, i.e., $dT_{\rm N}(B\neq 0)/dp$ is only small. In contrast, \ts($B$) shows clear volume changes, i.e., $dT_{\rm s}(B)/dp \neq 0$, while $dT_{\rm s}(B)/dB$ is very small. We conclude that, by applying external magnetic fields, the triple spin-structure-dielectric ordering phenomenon at \tn (B=0) is separated. The small magnetic field dependence of \ts\ and the absence of associated clear magnetization changes especially for fields above 5~T reveals the intrinsic structural/dielectric nature of this transition. Note, that the absence of clear changes in $M$ agrees to the observed small magnetic field dependence of the phase boundary \ts ($B$). The very small but finite positive slope $d$\ts $/dB > 0$ however shows that, at B $\leq$ 5 T, the magnetization increases at \ts ($B$) upon cooling while the opposite holds for $B\geq 6$~T where $d$\ts $/dB < 0$.

 In addition to the splitting of \tn\ into two anomalies at $B\neq 0$, there is also a kink in the thermal expansion coefficient at around \tp\ = 25~K (see Fig.~\ref{fig:LvsT}). \tp\ is nearly independent of the external magnetic field and it is neither associated with clear signatures in the magnetization nor in the dielectric properties~\cite{SannigrahiPRB2015} so that its nature remains unclear. The associated temperatures in the magnetic phase diagram (Fig.~\ref{fig:PD}) are marked by a grey line.

\subsection{Magnetostriction}

\begin{figure}
\includegraphics[width=1.0\columnwidth,clip] {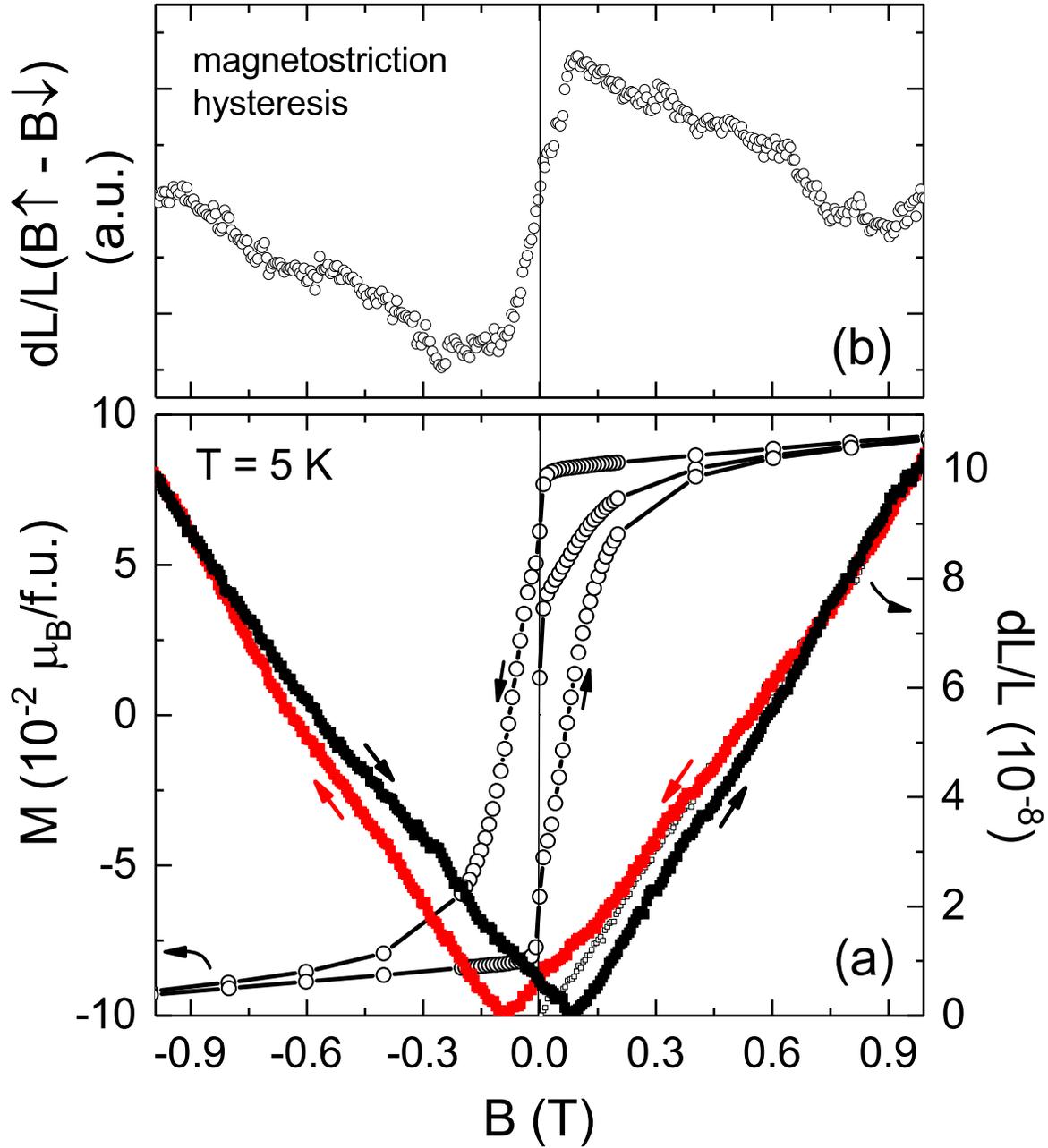}
\caption{(a) Hysteresis of the magnetization and of the relative length changes, and (b) the size of the magnetostriction hysteresis, i.e., $dL(B\uparrow)/L - dL(B\downarrow)/L$, in magnetic fields -1~T $\leq B\leq$ 1~T, at $T=5$~K.}\label{fig:hyst}
\end{figure}

\begin{figure}
\includegraphics[width=1.0\columnwidth,clip] {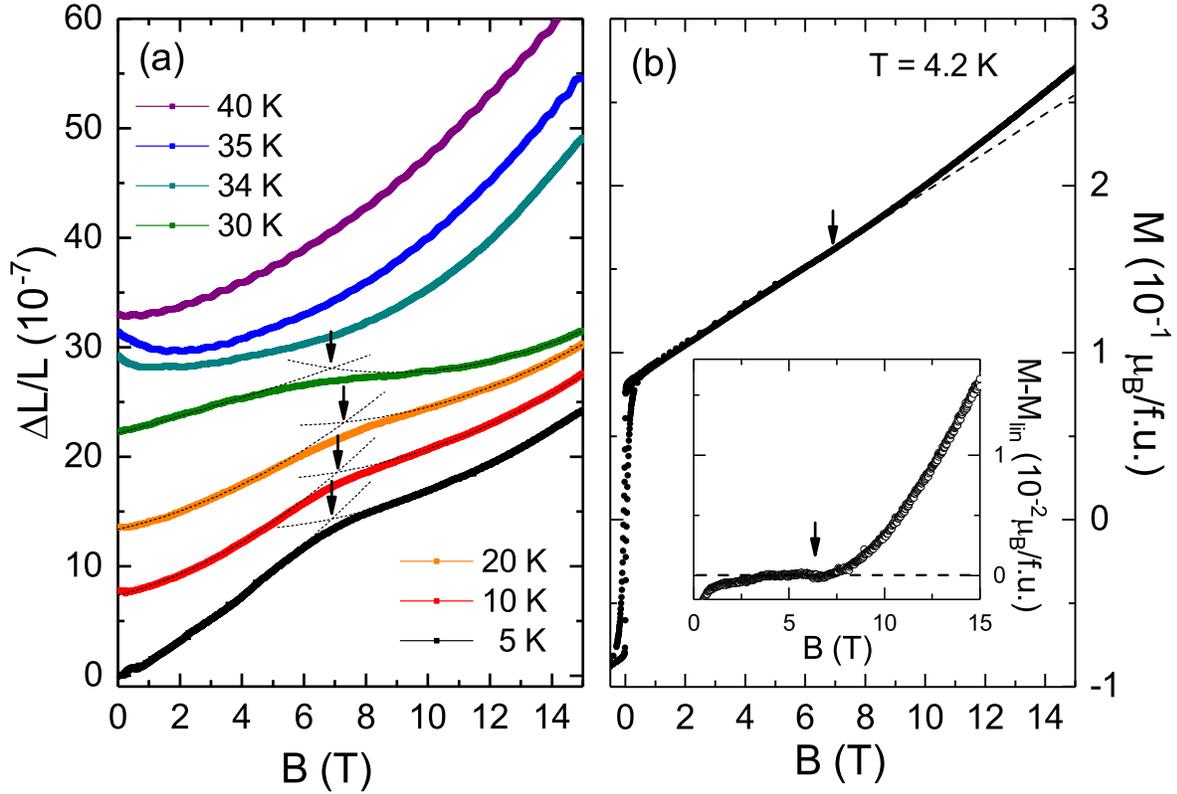}
\caption{(a) Relative length changes vs. external magnetic field, at various temperatures from $5$ to $40$~K. Arrows mark the critical field \bc . (b) Magnetization, at $T=4.2$~K, and the nonlinear behaviour obtained by subtracting the linear magnetization $M_{\rm lin}=M(B<6~{\rm T})$ (inset). Arrows in (b) show \bc\ from (a).}\label{fig:LvsB}
\end{figure}

In order to further investigate magnetoelastic effects in \cu , the longitudinal magnetostriction is considered. Fig.~\ref{fig:hyst} shows the magnetostriction at small magnetic fields while Fig.~\ref{fig:LvsB} displays the field dependence of the length in magnetic fields up to $B=15$~T. The hysteresis of the longitudinal magnetostriction at small fields resembles the hysteresis of the magnetization which is overlayed the magnetostriction data in Fig.~\ref{fig:hyst}a. Both quantities displays a clear hysteretic behaviour in which remanent values of the length $\Delta L/L = 8.8(2)\cdot 10^{-9}$ and magnetization ($M_{\rm R}(5~K)=0.082(1)$~\mbfu ) are obverved. The hysteresis region in the magnetostriction follows the one in the $M$ vs. $B$ curve.

In contrast to the low-field response which is affected by domain effects and illustrates properties of the canted antiferromagnetic low-field phase, the field induced length changes in the magnetic field range up to $B=15$~T presented in Fig.~\ref{fig:LvsB} enable to further complete the magnetic phase diagram. At $T=5$~K, the overall behaviour changes at about 7~T. In order to determine the associated phase boundary \bc ($T$), we have fitted the data well below and above this feature by polynomials (see the dashed lines in Fig.~\ref{fig:LvsB}a) and derived $B_{\rm C}$ from their intersection points. In the magnetisation at $T=4.2$~K (see Fig.~\ref{fig:LvsB}b), the feature in $dL(B)$ is associated with a vague kink marking the increase of susceptibility $\partial M/\partial B$. This is highlighted in the inset of Fig.~\ref{fig:LvsB}b where a linear contribution to $M$ extrapolated from the data at 1~T~$\leq B\leq$~5~T has been subtracted from $M$.


\subsection{Gr\"{u}neisen scaling}

\begin{figure}
\includegraphics[width=1.0\columnwidth,clip] {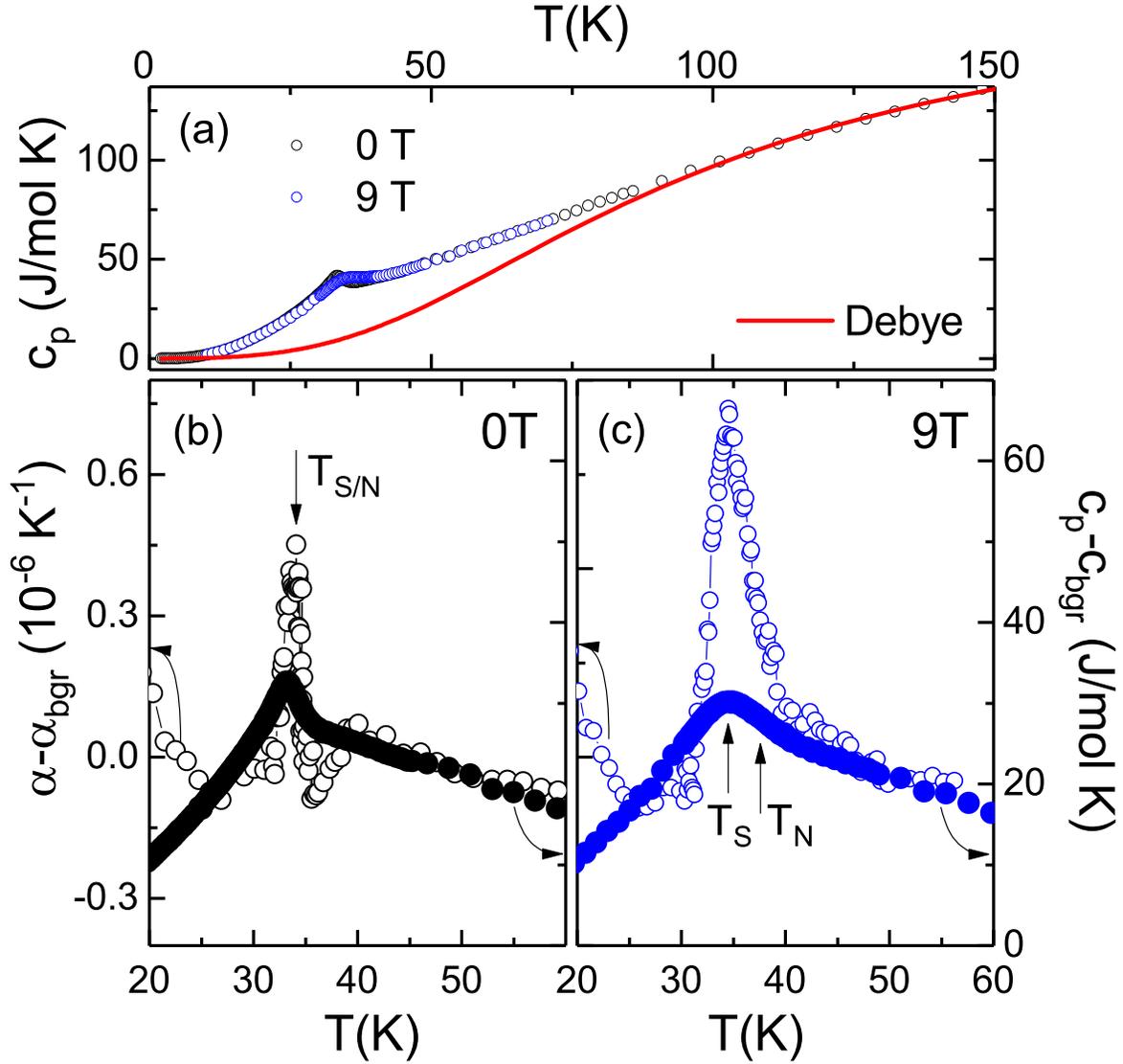}
\caption{(a) Specific heat \cp\ of \cu\ at $B=0$~T and 9~T. The line shows the Debye fitting of the background (see the text). (b,c) Gr\"{u}neisen scaling of the anomalous contributions in \al\ (open circles) and \cp\ (filled circles) for $B=0$ and 9~T. }\label{fig:Grueneisen}
\end{figure}

Comparing the anomalous contributions to the specific heat and to the thermal expansion enables further conclusions on the nature of the associated ordering phenomena. In a phenomenological approach, we have estimated the phonon contribution to the specific heat by means of the Debye function such that the entropy changes at high temperatures are described by the acoustic phonon background. This procedure yields $\Theta_{\rm D}\approx 387$~K and it allows an estimate of the background specific heat \cp $^{\rm bgr}$. Our procedure suggests anomalous entropy changes well above \tn , i.e., up to around 80~K. Interestingly, this temperature regime coincides with the region where the dielectric permittivity $\epsilon'$ changes in magnetic field.~\cite{SannigrahiPRB2015} Quantitatively, integrating the remaining specific heat changes (\cp -\cp$^{\rm bgr}$)/$T$, which are obtained by subtracting the result of the Debye fitting from the experimental data, yields about 30~\jmk . This value strongly exceeds the pure magnetic entropy $\Delta S^{\rm magn}=2R\ln (2)\approx 11.5$~\jmk . The large value agrees to the fact that the long-range ordered phase is of multiferroic nature, i.e., it includes spin, charge, and structural degrees of freedom which contribute significantly to the entropy changes. This is also demonstrated by the fact that the low-temperature anomalous specific heat does not follow a simple polynomial behavior, i.e., $T^n$ with $n \leq 3$, which is expected for pure magnetic order.


In Fig.~\ref{fig:Grueneisen}b and c the corrected specific heat and the corrected thermal expansion coefficient from Fig.~\ref{fig:1T}c, both obtained at $B=0$~T and $B=9$~T, are shown with appropriate scaling. At $B=9$~T, the specific heat anomaly is slightly broadened and entropy is shifted to higher temperatures as thermodynamically expected for the transition from a paramagnetic to a canted antiferromagnetic phase. For both fields, however, the temperature dependencies of \cp\ and \al\ are similar only above $T^*\approx 40$~K but differ around \tn\ (and \ts ) and below. According to the Gr\"{u}neisen law, such behaviour is expected in the absence of a single dominant energy scale which further emphasizes that several different degrees of freedom are driving the ordered state.

\subsection{High-frequency Electron Spin Resonance}

\begin{figure}
\includegraphics[width=1.0\columnwidth,clip] {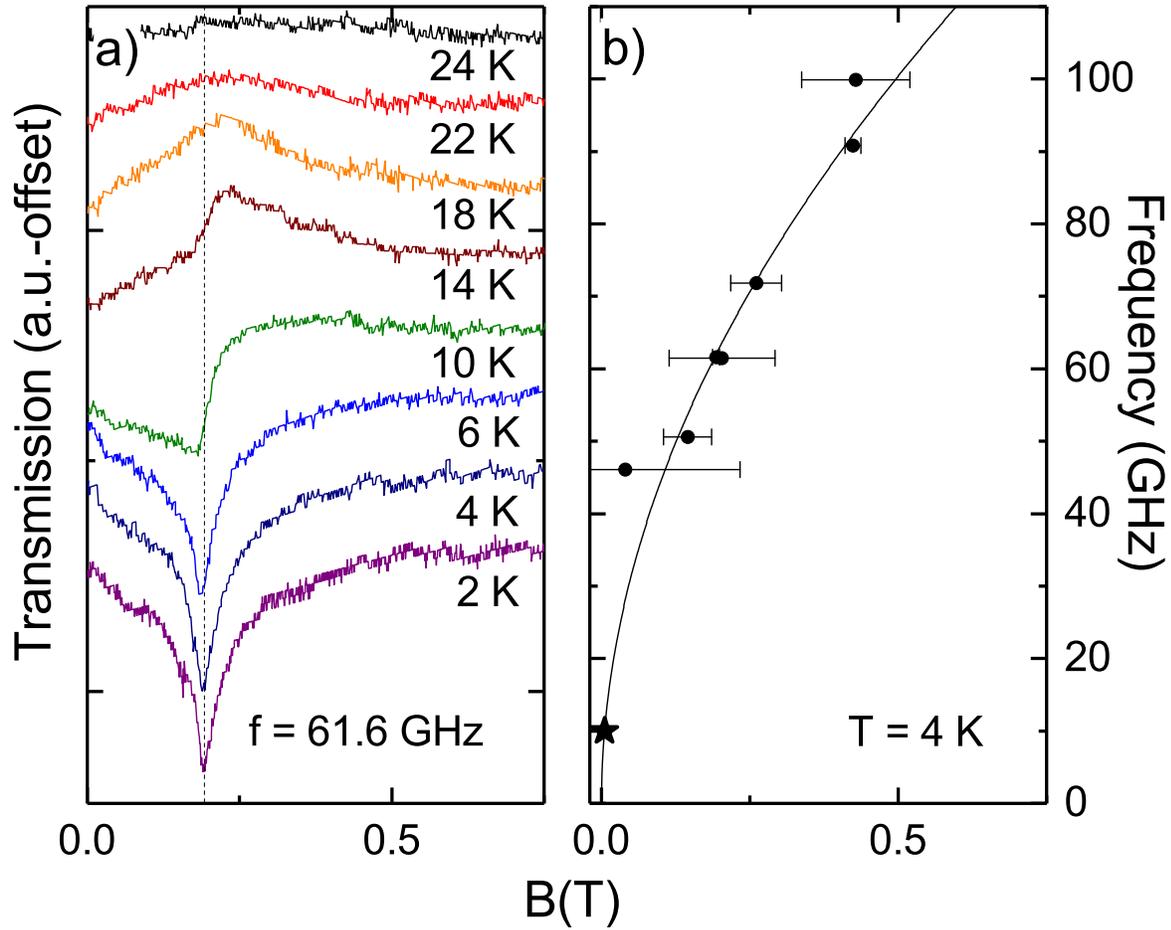}
\caption{(a) HF-ESR transmission spectra at $f = 61.6$~GHz in the temperature range 2~$\leq T\leq$~24~K. (b) Resonance frequencies vs. magnetic field at $T=4$~K. The solid line is a fit according to equation~\ref{eq:ESRFM}. The star shows the X-band ESR resonance from Ref.~\cite{pommer2003interplay}.}\label{fig:ESR}
\end{figure}

While HF-ESR measurements are susceptible to collective $q=0$ spin excitations in the long range spin-ordered state, i.e. antiferromagnetc resonance (AFMR) modes, the large AFM gap of $\sim 10$~meV inferred from inelastic neutron scattering on \cu~\cite{banerjee2016spin} rules out the observation of AFMR modes at frequencies below 2~THz. However, the ESR spectra taken at $f=61.6$~GHz  shown in Fig.~\ref{fig:ESR}a display a clear resonance peak appearing at $T<20$~K and in low magnetic fields.~\footnote{In addition to this FMR mode, a powder broadened paramagnetic signal is observed at higher magnetic fields. Its integrated intensity follows a Curie-law and the $g$-factor amounts to $g_\perp = 2.06$ and $g_{||} = 2.33$ which is typical for Cu$^{2+}$-spins in octahedral environment. We hence attribute this resonance feature to a small amount of impurity spins.} We attribute this resonance to a ferromagnetic resonance (FMR) branch associated with the canted magnetic moment induced by DM interactions. At $T=2$~K, the peak has a Lorentzian shape with a center field of \bres\ = 0.19~T and peak width of $\Delta B = 0.06$~T. Upon heating, the intensity of the resonance feature decreases and the peak broadens. The position of the peak, i.e. the resonance fields \bres , does not change in the temperature range under study.


Measurements of the ESR signal at $4$~K and various frequencies allow to construct the frequency-magnetic field phase diagram of the resonance. As shown in Fig.~\ref{fig:ESR}b, the resonance field slightly shifts to higher frequencies with increasing magnetic field. Above $100$~GHz, the resonances cannot be detected anymore. Note, that in a previous X-band ESR study on \cu\ a resonance feature was observed at $B_{\rm res}=0.01$~T which is also shown in Fig.~\ref{fig:ESR}b.~\cite{pommer2003interplay} Resonances of the DM moments can be described by means of a standard phenomenological treatment of ferromagnetic resonance. Motivated by a recent neutron study\cite{gitgeatpong2017nonreciprocal}, a two-sublattice antiferromagnetic resonance model with in-plane-type anisotropy is applied, which includes DM interactions causing spin canting.~\cite{bahr2014low,Pincus1960} In this model, two antiferromagnetic resonance (AFMR) branches as well as one ferromagnetic resonance (FMR) mode appear.~\footnote{Note, that the FMR mode for $B\|c$-axis is supposed to be field independent and gapless so that it does not show up in the spectra.} Due to the large AFM gap, we only consider the FMR mode which, for the magnetic field $B$ being applied in the $bc$-plane, i.e., the anisotropy plane, is given by \cite{bahr2014low}

\begin{equation}
\omega_{\rm FM} = \sqrt{(g_{\rm bc}\mu_{\rm B}B)^2+4(2J_{\rm eff}+\tilde{D})M_{\rm FM}B}.
\label{eq:ESRFM}
\end{equation}

Here, $J_{\rm eff}$ is the effective isotropic exchange, $M_{\rm FM} = 0.082(1)\,\mu_{\rm B} $ the in-plane ferromagnetic moment (see Fig.~\ref{fig:hyst}), $\tilde{D}$ the effective in-plane anisotropy, and $g_{\rm bc}$ the $g$-factor in the $bc$-plane. Applying the constraint for the AFMR gap $\sqrt{32 J_{\rm eff}\tilde{D}S^2} = 10$~meV as detected in a recent neutron study on a powder sample from the same batch as studied here~\cite{banerjee2016spin}, and the $g$-value $g_{\rm bc} \approx 2$, we obtain $J_{\rm eff} = 8(3)$~meV and $\tilde{D} = 1.6(1)$~meV.

The obtained value of the effective two-sublattice antiferromagnetic exchange constant is consistent with the dominant third-nearest-neighbour exchange interaction $J_3$ inferred from inelastic neutron data which in addition to slightly smaller nearest and next-nearest-neighbour couplings $J_1$ and $J_2$ governs the long range spin ordered phase.~\cite{banerjee2016spin} Note, that the in-plane $g_{\rm bc}$-factor cannot be determined more precisely because the slope of the resonance branch is dominated by the DM interaction. From the temperature independence of the resonance field up to 24~K we conclude that the effective DM-field does not change with temperature.~\cite{turov1965physical}

\section{Discussion}

Our data imply strong coupling between the structure and the magnetic and dielectric properties in \cu . This shows up, i.e., in a pronounced peak-like anomaly in \al\ at \tn . The weak first order character of this transition is confirmed by a small temperature hysteresis of the magnetization at \tn\ (see Ref.~\cite{SannigrahiPRB2015}). Failure of Gr\"{u}neisen scaling well above \tn\ implies that there are at least two ordering phenomena of similar relevance. Indeed, at the temperature $T^*$ below which Gr\"{u}neisen scaling fails, ferroelectric polarization starts to evolve.~\cite{SannigrahiPRB2015} We conclude that both spin and dielectric degrees of freedom are driving the ordering process. This conclusion of multiple dominant phenomena is supported by the magnitude of the measured anomalous entropy changes which are observed exactly in the same temperature regime where the dielectric permittivity is affected by external magnetic fields. The total anomalous entropy changes are more than twice of the spin entropy which confirms that additional, i.e., dielectric and structural, degrees of freedom accompanying spin ordering are associated with significant entropy changes of similar magnitude.

Due to pronounced coupling of the magnetic and the dielectric properties to the structure, thermal expansion studies enable constructing the magnetic phase diagram. Despite several similarities, the magnetic phase diagram in Fig.~\ref{fig:PD} displays clear differences as compared to data which have been recently reported in Ref.~\cite{gitgeatpong2017}. E.g., the observed anomalies at $B>5$~T in Fig.~\ref{fig:MvsT} disagree to any anomaly reported in the previously published phase diagram. Note, that differences might result from the different V-O-V bond angles~\cite{Bhowal2017} in the polycrystalline samples studied at hand and the single crystalline one in Ref.~\cite{gitgeatpong2017}, respectively. The magnetostriction data indicate a transition at \bc\ which phase boundary, at low temperatures, is similar to the metamagnetic transition observed in Ref.~\cite{gitgeatpong2017}.  Though it might be tempting to assign the findings at hand to those in Ref.~\cite{gitgeatpong2017}, we note that both the temperature dependence of the phase boundary \bc ($T$) as well as the anomalies in the magnetisation and the magnetostriction do not agree to what is reported in Ref.~\cite{gitgeatpong2017}. To be specific, our data imply no significant temperature dependence of \bc\ which is inferred from kinks in the magnetization and magnetostriction. In contrast, in Ref.~\cite{gitgeatpong2017}, the phase boundary at \bc\ is of discontinuous spin-flop-like nature but it does not extend to \tn\ and shows a strong temperature dependence. In zero field, in addition to the coupled magnetic/dielectric/structural transition at \tn\ we observe a change in the thermal expansion coefficient at \tp\ $\approx 25$~K. This feature is not associated with significant magnetization changes which somehow agrees to the small slope $dT{'}/dB$.

The intimate coupling of the spontaneous magnetic and electric moments to the structure is particularly evident if the magnetostriction in the hysteresis region -1 T $< B <$ 1 T is considered. The fact that magnetic hysteresis is not only associated with ferroelectric hysteresis but also with structural distortion supports the suggested magnetostrictive nature of the giant ferroelectric polarization in \cu .~\cite{SannigrahiPRB2015} Comparison of the magnetostriction loop in Fig.~\ref{fig:hyst} with the hysteresis of the magnetization implies that magnetic field driven switching of the length is directly associated with the magnetic domain structure.

Despite the large antiferromagnetic gap in \cu , the ferromagnetic resonance branch is detected by means of HF-ESR measurements. This in-gap excitation branch is associated with the canted magnetic moment arising from Dzyaloshinsii-Moriya interaction and it enables quantitative estimates of the DM-parameter and the effective exchange constant. The obtained value of the effective exchange constant $J_{\rm eff} = 8(3)$~meV is larger than the nearest neighbour exchange constants determined by neutron scattering of \cu\ powder $J_1 = 4.67$~meV and $J_2 = -0.8$~meV, but in good agreement with the dominant exchange $J_3 = 9$~meV.~\cite{banerjee2016spin} Although a similar Hamiltonian was applied for the data analysis, the anisotropy parameter obtained by inelastic neutron scattering (INS) of a single crystal sample considerably differs from our results while $J_{\rm eff}$ is similar.~\cite{gitgeatpong2017nonreciprocal} To be specific, the isotropic exchange constants from the single crystal INS amount to $J_1 = 2.67$~meV, $J_2 = 2.99$~meV, and $J_3 = 5.42$~meV. With the DM-parameter $D = 2.79$~meV, these parameters sum up to  $J_{\rm eff}^{\rm INS} =\sqrt{(J_1+J_2+J_3)^2+D^2} = 11.4$~meV. The effective anisotropy from Ref.~\cite{gitgeatpong2017nonreciprocal}, i.e., $\tilde{D}^{\rm INS} = J_{\rm eff}^{\rm INS}-(J_1+J_2+J_3)-2G = -0.2$~meV, where $G$ is the anisotropic exchange interaction, strongly differs from $\tilde{D} = 1.6(1)$~meV obtained from the analysis the HF-ESR data at hand.~\footnote{Note again the differences in the V-O-V bond angles in the single crystal and the polycrystalline sample studied here.~\cite{Bhowal2017}}

\section{Summary}

We have investigated magnetoelectric coupling and low-energy magnetic excitations in multiferroic \cu\ by detailed thermal expansion, longitudinal magnetostriction, specific heat, magnetization, and HF-ESR measurements in magnetic fields up to 15~T. The resulting magnetic phase diagram differs from a previously reported one. Dichotomy between the field effect on the magnetization and the thermal expansion indicates the effects of magnetic fields $B\|c$ and $B\perp c$, respectively, on the polycrystalline sample. By applying external magnetic fields, the triple spin-structure-dielectric ordering phenomena at \tn ($B=0$) are separated. At $B\neq 0$, the evolution of the spontaneous magnetic moment at \tn ($B$) is not associated with significant structural changes, i.e., the anomaly temperature is rather pressure independent and the transition may be considered predominately magnetic. On the other hand, the thermal expansion anomaly at \ts (B$\perp$c) reveals the intrinsic structural/dielectric nature of this transition by the absence of associated clear magnetization changes. Well above \tn , we find anomalous entropy changes in the temperature regime where anomalous dielectric, magnetic and structural response is detected. Their magnitude as well as failure of Gr\"{u}neisen scaling suggests that magnetic, structure and charge degrees of freedom are driving multiferroic order concomitantly. In addition, our magnetostriction data support an exchange-striction driven mechanism of ferroelectricity. Despite the large AFM gap, we observe low-energy magnetic in-gap excitations in the spin ordered phase which are associated with the canted magnetic moment arising from Dzyaloshinsii-Moriya interaction. The anisotropy parameter $\tilde{D}=1.6(1)$~meV indicates a sizeable ratio of DM-exchange and isotropic magnetic exchange.

\section{Acknowledgements}
RK gratefully acknowledges fellowship of the Marsilius Kolleg Heidelberg. JW acknowledges support from the IMPRS-QD. RK and MAH acknowledge the Federal Ministry of Education and Research (BMBF) and the German-Egypt Research Fund GERF IV (project 01DH17036).


\end{document}